\documentclass[useAMS,usenatbib]{mn2e}

\def\inpdir{.}
\usepackage{\inpdir/psfig}
\usepackage{graphicx}
\input \inpdir/abbreviations.in
\renewcommand{\epsilon}{\varepsilon}
\renewcommand{\phi}{\varphi}
\title[The evolution of binary star clusters]{The evolution of binary
star clusters and the nature of NGC2136/NGC2137}

\author[S. F. Portegies Zwart$^{1,2}$,
        S. P. Rusli$^{3}$  
        ]
       { 
        S. F. Portegies Zwart$^{1,2}$\thanks{spz@science.uva.nl}
        S. P. Rusli$^{3}$\\  
	$^{1}$Astronomical Institute 'Anton Pannekoek', 
	      University of Amsterdam, Amsterdam, the Netherlands\\
	$^{2}$Computational Science Section, University of Amsterdam, 
	      Amsterdam, the Netherlands\\
	$^{3}$Astronomy Study Program and Bosscha Observatory, 
            Faculty of Mathematics and Natural Sciences, \\
            Institute of Technology Bandung, Bandung, Indonesia\\
       }
\begin{document}
\date{Accepted 2006 ???. Received 2006 ???; in original form 2006 ???}
\pagerange{\pageref{firstpage}--\pageref{lastpage}} \pubyear{2006}
\maketitle
\label{firstpage}
\begin{abstract}

We study the evolution of bound pairs of star clusters by means of
direct N-body simulations. Our simulations include mass loss by
stellar evolution.  The initial conditions are selected to mimic the
observed binary star cluster NGC\,2136 and NGC\,2137 in the Large
Magellanic Cloud.  Based on the rather old ages ($\sim 100$\,Myr),
masses, sizes of the two clusters and their projected separation, we
conclude that the cluster pair must have been born with an initial
separation of 15--20\,pc.  Clusters with a smaller initial separation
tend to merge in $\aplt 60$\,Myr due to loss of angular momentum from
escaping stars.  Clusters with a larger initial separation tend to
become even more widely separated due to mass loss from the evolving
stellar populations.
The early orbital evolution of a binary cluster is governed by 
mass loss from the evolving stellar population and by loss of angular
momentum from escaping stars. Mass loss by stellar winds and
supernovae explosions in the first $\sim 30$\,Myr causes the binary to
expand and the orbit to become eccentric.
The initially less massive cluster expands more quickly than the
binary separation increases, and is therefore bound to initiate mass
transfer to the more massive cluster.  This process is quite contrary
to stellar binaries in which the more massive star tends to initiate
mass transfer.  Since mass transfer proceeds on a thermal timescale
from the less massive to the more massive cluster, this semi-detached
phase is quite stable, even in an eccentric orbit until the orbital
separation reaches the gyration radius of the two clusters, at which
point both clusters merge to one.

\end{abstract}

\section{Introduction}\label{sec:intro}

The large Magellanic cloud (LMC) is a rich environment with many
relatively young star clusters
\citep{2003MNRAS.338...85M,2003MNRAS.338..120M}. Several of these
clusters appear closer together on the sky than expected from
statistical arguments, i.e, they look like binary clusters
\citep{1991A&AS...87..335B,2002A&A...391..547D}.  The LMC contains a
total of 69 cluster pairs with a separation $R \aplt 1'.3$
\citep{1988MNRAS.230..215B} which corresponds to $\aplt 19$ pc by
adopting distance modulus of 18.5 mag to the LMC
\citep{2003A&A...410..887G}, many of which have similar colors
\citep{1988AJ.....96.1625K}.  The small Magellanic cloud also contains
a relatively large number of paired clusters
\citep{1990A&A...230...11H}.  Binary star clusters are expected to
merge on a time scale of a few times 10\,Myr
\citep{1988MNRAS.230..215B}, and NGC\,2214 may be an LMC cluster in
the process of merging \citep{1988A&A...203L...5B} (but see
\cite{1995MNRAS.274.1225B} for counter arguments).

Well known examples of such binary clusters are SL\,538/NGC\,2006 (SL
537) \citep{1998A&A...339..773D}, NGC\,1850
\citep{1993AJ....105..938F} and NGC\,2136 (SL\,762, at $\alpha=$
5$^{\rm h}$\, 53$^{\rm m}$\, 17$^{\rm s}$; $\delta= -69^\circ$\,
32'')/NGC\,2137 \citep{1995A&A...299L..37H}.  Confirmation of the
binarity of such apparent pairs of star clusters should come from
detailed measurements of their orbital parameters, rather than
statistical arguments. However, with an orbital period of tens of
million years such measurements are not trivial.  In the Antennae
galaxies \citep{2005AJ....130.2104W,2005ApJ...631L.133F} and in the
young starburst galaxy M51
\citep{2000MNRAS.319..893L,2005A&A...443...79B}, star clusters appear
to be formed in groups often consisting more than two.  Apparently,
star clusters in general are born in conglomerates in which several
clumps tend to form within a relatively small time interval presumably
initiated by galaxy-galaxy interactions. The individual star clusters
in such conglomerates may subsequently merge in due time
\citep{2002Ap&SS.281..355F,2002CeMDA..82..113F}.  Cluster binaries may
form in a comparable environments, in which case their ages have to be
at least comparable.  Alternatively, a more massive cluster could
possibly captures a lower mass cluster in a tidal event, resulting in
two cluster with different ages.  However, this process is likely to
result in a single cluster rather than a binary
\citep{1999IAUS..190..443D}.

In this study we mainly concentrate on the two star clusters NGC\,2136
and NGC\,2137, which attract attention because the projected distance
between them is only about 20\,pc (1'.34)
\citep{1994AGAb...10..198S}. In addition, both clusters have similar
ages of about $100\pm20$\,Myr \citep{2000A&A...360..133D} and
identical metalicity \citep{1995A&A...299L..37H}. With a primary mass
of 26300--28200\,\msun\, \citep{2003MNRAS.338..120M} and a mass ratio
$q \simeq 0.17$, their orbital period is about 46\,Myr. If bound, both
clusters have orbited each-other about twice within their lifetime.
Regretfully it is currently not known whether or not the clusters
really form a bound pair, but at the moment this seems to be the most
logical conclusion.

In this study we explore the initial conditions for which a pair of
star clusters could survive until an age of about 100\,Myr. We focus
in particular on the observed pair NGC\,2136 and NGC\,2137
\citep{1995A&A...299L..37H}, as they form an observational motivation
for this exercise.  Assuming that the two clusters form a bound pair
we are able to limit the initial cluster parameters. We identify two
main regimes in binary cluster evolution. Small initial orbital
separation ($\aplt 12$\,pc) tends to lead to a merger on a time scale
of less than about 60\,Myr, whereas large initial separation ($\apgt
17$\,pc) causes the two star clusters to recede. Clusters with
initially a larger separation become very vulnerable to disruption by
passing other star clusters, giant molecular clounds or the background
potential of the host galaxy.  Between about 12\,pc and 17\,pc is a
range of orbital separations between which the period changes little
and in which the cluster binary is able to survive for more than
100\,Myr. In these cases the change in orbital angular momentum by
stellar mass loss and escapers is compensated by the redistribution of
internal angular momentum in the rotation of the clusters.

\section{Validation and analytic considerations}\label{sec:analytic}

We perform direct $N$-body simulations of binary star clusters using
the kira integrator of the starlab simulation environment
\citep{2001MNRAS.321..199P}. This $N$-body code computes
inter-particle forces by direct summation and the integration of the
equations of motion is carried out with a fourth-order Hermite
predictor-corrector scheme \citep{1992PASJ...44..141M} with block time
steps \citep{1986ApJ...307..126M}. The greatest speed is obtained with
the GRAPE-6 special purpose computer
\citep{1997ApJ...480..432M,2003PASJ...55.1163M,2004PASJ...56..521M,1998sssp.book.....M}
and we use the MoDeStA\footnote{see {\tt
http://modesta.science.uva.nl}} platform in Amsterdam. Some of our
calculations incorporate the evolution of single stars and binaries
via the {\tt SeBa} package \citep{1996A&A...309..179P}. However, since
our simulations all started with single stars, binaries play a minor
role as only a few are formed during the course of the simulations by
3-body dynamical capture \citep{1976A&A....53..259A}.

Following the stellar evolution package {\tt SeBa}, stars more massive
than about 25\,\msun\, turn into black holes in a supernova explosion,
and stars more than about 8\,\msun\, turn into neutron stars. The
latter receive a kick velocity upon formation from the
Paczinsky-Hartman distribution with a dispersion of 300\,\kms
\citep{1997A&A...322..127H}, black holes of mass $m_{\rm bh}$ receive
kicks from the same distribution but with a dispersions of $300 \times
1.4\msun/m_{\rm bh}$\,\kms\, (see \cite{1998A&A...332..173P} for
details). During the simulation we keep track of the individual
clusters by using a clump-finding algorithm
\citep{1998ApJ...498..137E}. Masses referring to the individual
clusters when discussing the results in Tab.\,\ref{Tab:Results} are
therefore an over estimate as the potential of a possible companion
cluster is ignored in this procedure.

\subsection{Binary-cluster evolution without stellar evolution}
\label{Sect:Makino}

To test the merger time for binary clusters we started by reproducing
the calculations of \citep{1989PASJ...41.1117S,1991Ap&SS.185...63M}.
They performed for that time large scale N-body simulations using a
tree code \citep{1986Natur.324..446B} with a softening of $\epsilon =
0.025\rvir$. Here \rvir\, is the initial virial radius of the primary
cluster. All stars in their models had the same mass and they ignored
stellar evolution.  The initial density profile for all simulations
was a \cite{1966AJ.....71...64K} model with $W_0 = 7$, the tidal
radius is then $r_{\rm tide} \simeq 7\rvir$. When on a circular orbit
with separation $a=10$\,\rvir\, both clusters are initially in
Roche-lobe contact. We summarize the initial conditions and results in
Tab.\,\ref{Tab:Makinoetal}.  These simulations are carried out in
dimension less $N$-body units \citep{1986LNP...267..233H} in which $M
= G = \rvir = 1$. Here $M$ is the total cluster mass and $G$ is Newton's
constant.

\begin{table}
\caption[]{Overview of the initial conditions explored by
\cite{1989PASJ...41.1117S,1991Ap&SS.185...63M} and reproduced here.
In the first column is the name of the simulation followed by the
number of particles in two clusters. The subsequent two columns give
the initial separation with which the clusters were born ($R_i$) and
their initial orbital eccentricity ($e_i$). Note that cluster B was
born near semi-latus rectum. The last four columns give the results of
the simulations.  These give the time of merger in standard $N$-body
units \citep{1986LNP...267..233H} ($t_f$), and the total number of
stars in the merger cluster as fraction of the total number of stars
in the initial cluster $f_N = (N_f+n_f)/(N_i+n_i)$.  The last two
columns give these numbers for the simulations without softening.  }
\bigskip
\begin{tabular}{lrrrrrrrr}
Model&  $N$ & $n$ & $R_i$& $e_i$& $t_f$ & $f_N$
                                           & $t_f$ & $f_N$ \\
     &      &     & [$\rvir$]    &      
                           & \multicolumn{2}{c}{$\epsilon = 0.025\rvir$} 
                           & \multicolumn{2}{c}{$\epsilon = 0$} \\
\hline
A         &  2048& 2048&         10 &   0.0& 370 & 0.95& 270 & 0.93 \\
B         &  2048& 2048&          6 &   0.5& 190 & 0.96& 150 & 0.95 \\
C         &  4096& 2048&          6 &   0.0& 105 & 0.97& 100 & 0.96 \\
D         &  4096& 2048&          6 &   0.0& 145 & 0.97& 120 & 0.96 \\
\hline
\label{Tab:Makinoetal}
\end{tabular}
\end{table}

When adopting the same initial conditions as
\cite{1989PASJ...41.1117S} and \cite{1991Ap&SS.185...63M} we
reproduced their results. However, when adopting zero softening
$\epsilon = 0$ our clusters tended to merge somewhat earlier and the
merger product was somewhat less massive (see
Tab\,\ref{Tab:Makinoetal} for details).  This is not unexpected as
softening in the simulations has the effect of reducing close
encounters causing the relaxation time for the cluster pair to
increase. As a result, softening tends to reduce mass loss, by reducing
the number of close encounters, and increases the merger time.

\subsection{Analytical considerations}\label{Sect:analytic}

The relation between mass loss and angular momentum loss from the
binary can be computed from the conservative assumption that both
clusters are in synchroneous rotation.  Since in our simulations the
clusters are initially not rotating we then overestimate the rate of
angular momentum loss.

The angular momentum of two point masses in a circular orbit is
\begin{equation}
  J =      \mu a^2 \omega 
    \equiv
           {q \over (q+1)^2} a^2 \omega M.
\label{Eq:J}
\end{equation}
Here $\mu = m_pm_s/M$, with $M \equiv m_p+m_s$ and $q \equiv
m_s/m_p$.  The full differential can then be written as
\begin{equation}
  D\ln J = d\ln \mu + 2d\ln a + d\ln \omega.
\label{Eq:DlnJ}
\end{equation}
With Kepler's \cite{Kepler:1609} third law ($a^3 \propto M/\omega^2$) 
\begin{equation}
  3d\ln a = d\ln M - 2d\ln \omega,
\end{equation}
we cen reduce Eq.\,\ref{Eq:DlnJ} by eliminating the orbital velocity
\begin{equation}
  d\ln J = d\ln m_p + d\ln m_s - {1 \over 2} d\ln M + {1 \over 2} d\ln a.
\label{Eq:dlnJ}
\end{equation}
With the assumption that mass is lost from the binary with angular momentum 
\begin{equation}
   dJ/dM = \gamma J/M
\end{equation}
Eq.\,\ref{Eq:dlnJ} reduces to
\begin{equation}
  d\ln a = (2\gamma + 1) d\ln M - 2 d\ln m_p - 2 d\ln m_s.
\end{equation}
Which, after integrating from the initial conditions to the final
conditions results in
\begin{equation}
 {a_f \over a_i} = \left(
                    {M_f \over M_i}
                   \right)^{2\gamma + 1}
                   \left(
                    {m_{p,f}m_{s,f} \over m_{p,i}m_{s,i}}
                   \right)^{-2}.
\label{Eq:afai}
\end{equation}

If mass is predominantly lost by stellar evolution and if both
clusters have the same initial mass function, we may adopt that the
mass ratio $q$ remains constant throughout the mass loss process.
In this case Eq.\,\ref{Eq:DlnJ} reduces to
\begin{equation}
  d\ln J = d\ln M + 2d\ln a + d \ln \omega,
\end{equation}
which after making the same assumption about the amount of angular
momentum lost per unit mass reduces to
\begin{equation}
  d\ln a = (2\gamma - 3) d \ln M.
\end{equation}
Integrating between the initial and final parameters then gives
\begin{equation}
   {a_f \over a_i} = \left(
                     {M_f \over M_i}
		     \right)^{2\gamma - 3}.
\label{Eq:af_ai}
\end{equation}
If mass is lost adiabatically and isotropically (i.e, $\gamma = 1$)
Eq.\,\ref{Eq:af_ai} reduces to the classical result in which $aM =$
constant.

\subsubsection{Results for the analytic prescription}\label{Sect:toymodel}

With the above prescription we can calculate the orbital evolution of
a binary star clusters in the absence of dynamical friction.  In the
presence of stellar evolution most mass is likely to be lost via the
evolving stellar population. 

\begin{figure}
\psfig{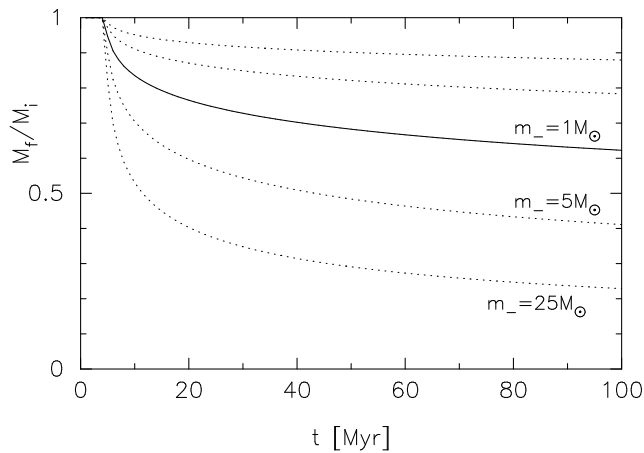}
   \caption[]{
     The mass of a star cluster as a function of time.  From top to
     bottom the varous curves are calculated with a Salpeter initial
     mass function with a maximum mass of $m_+=100$\,\msun\, and with
     a minimum mass of $m_- = 0.04$\,\msun, 0.2, 1.0, 5 and $m_- =
     25$\,\msun.  We adopted Eq.\,\ref{Eq:tms} to calculate the
     lifetime of a star with mass $m_\star$ (see
     \S\,\ref{Sect:toymodel} for details).
\label{fig:M_analytic} }
\end{figure}

To calculate the evolution of the cluster mass we adopte an initial
mass function, for which we use a \cite{1955ApJ...121..161S} between a
minimum mass of $m_-$ and a maximum of $m_+$.  We now assume that all
stars with a mass $m_\star > m_{\rm to}$ are lost from the clusters.
The lifetime of a star of mass $m_{\star}$ is calculated using the fit to
detailed stellar evolution tracks by \citep{1989ApJ...347..998E}
\begin{equation}
     t_\star(m_\star) = {3600 + 940m_\star^{2.5} + 1.4m_\star^{4.5}
                      \over
                      0.033m_\star^{1.5} + 0.35m_\star^{4.5}}.
\label{Eq:tms}
\end{equation}
According to this methodology stellar remnants are lost from the
cluster. As long as the cluster is younger than about 40\,Myr this
assumption does not result in an overestimate of the mass loss since
black holes and neutron stars are likely to receive a velocity kick
upon formation. As a result the majority of compact objects are
ejected from the cluster in an early stage. At later time, when white
dwarfs form, the use of Eq.\,\ref{Eq:tms} tends to overestiemate the
mass loss, as white dwarfs tend to stay in the cluster.  In
fig.\,\ref{fig:M_analytic} we present the evolution of the cluster
mass for several choises of $m_-$.

\begin{figure}
\psfig{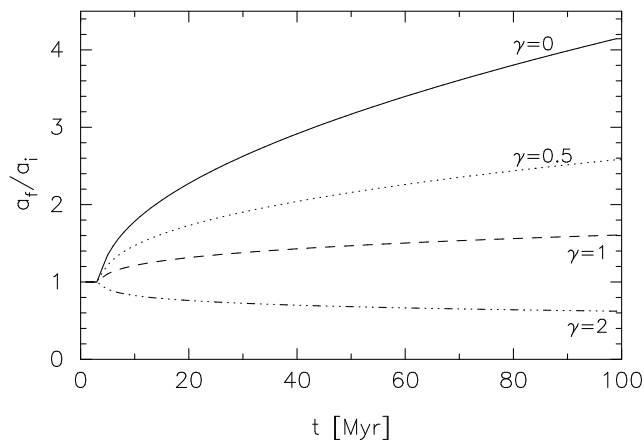}
   \caption[]{
     The variation in the orbital separation as a function of time for
     various choises of $\gamma = 0$, 1, 1.5 and $\gamma=2$ (see
     Eq.\,\ref{Eq:afai}).  Each calculations is performed with a
     Salpeter initial mass function between $m_-=1$\,\msun\, and $m_+
     =100$\,\msun.
\label{fig:a_analytic} }
\end{figure}

In fig.\,\ref{fig:a_analytic} we present the evolution of the orbital
separation under several assumptions for the amount of angular
momentum carried away per unit mass (see Eq.\,\ref{Eq:afai}).  These
calculations are performed under the assumption that the time scale
for mass loss is long compared to the orbital period. We will later
see that this is not a valid assumption in the first few tens of
million years.

\subsubsection{Results for two-body integration}\label{Sect:2body}

To test the above derivation we implement the mass loss prescription
adopted in \S\,\ref{Sect:toymodel} and integrate the equations of
motion of a two-body system.  During integrating the equations of
motion we allow the two point masses to lose mass.  Note than since
the prescription for stellar mass loss depends on physical scales we
provide a scaling to stellar units to the 2-body integration.

\begin{figure}
\psfig{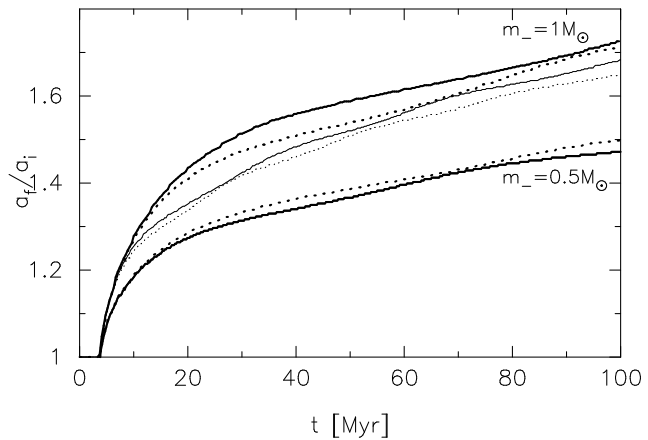}
   \caption[]{
     Semi-major axis vs. time for various initial conditions of two
    mass losing particles in an initialy circular orbit.  The top (4)
    curves are calculated with $m_-=1.0$\,\msun\, and the bottom
    curves with $m_-=0.5$\,\msun.  The thick curves are for $a_i =
    20$\,pc, the thin curves for $a_i=10$\,pc.  The solid curves are
    for $q=1$, and the dotted curves for $q=0.1$.
\label{fig:a_2body} }
\end{figure}

\begin{figure}
\psfig{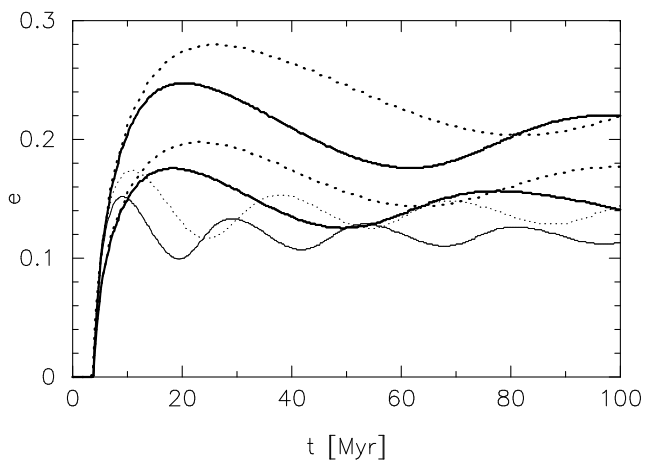}
   \caption[]{
    Evolution of the orbital eccentricity for the two orbiting
    clusters with various initial conditions of two mass losing
    particles in an initialy circular orbit (see
    \S\,\ref{Sect:2body}).  The top (4) curves are calculated with
    $a_i=20$\,pc and the bottom curves with $a_i=10$\,pc.  The thick
    curves are for $m_-=1.0$\,\msun\, and the thin curves for
    $m_-=0.5$\,\msun.  The solid curves are for $q=1$, and the dotted
    curves for $q=0.1$.
\label{fig:e_2body} }
\end{figure}

To enable a direct comparison with the simulations reported in
\S\,\ref{sec:simulation} we perform our calculations with similar
initial conditions.  In figs\,\ref{fig:a_2body} and
\,\ref{fig:e_2body} we present the evolution of the orbital parameters
($a$ and $e$, respectively) for two point masses. The initial orbit
was circular, and we adopted a primary mass of 27500\,\msun\, and a
mass ratio of $q=1.0$ and $q=0.16$.  

The mass loss prescription described in \S\,\ref{Sect:toymodel} is
then combined with the \cite{1955ApJ...121..161S} initial mass
function, from which we then can calculate the rate of mass loss.
Calculations are performed with a maximum mass of $m_= - 100$\,\msun\,
and with a minimum of $m_-=1.0$\,\msun\, and $m_-=0.5$\,\msun. The
evolution of the cluster mass for these parameters are presented in
Fig.\,\ref{fig:M_analytic}.

The orbital expansion (Fig.\,\ref{fig:a_2body}) is comparable to that
of the analytic prescription using $\gamma =1$ (see
Fig.\,\ref{fig:a_analytic}).  The semi-major axis for the choise of
$m_-=1$\,\msun\, increases more quickly than for $m_- = 0.5$\,\msun.
In this two-body approach, the friction between the clusters and
possible transfer of stars from one to the other are ignored.  As a
result, the initial mass ratio and the initial orbital separation have
little influence on the lifetime of the binary cluster.  For orbital
separations smaller than about 10\,pc, it will turn out that dynamical
friction has an important effect on the orbital evolution.  We will
quantify this in \S\,\ref{sec:simulation} where we perform the full
$N$-body simulation of the binary star cluster.

In Fig.\,\ref{fig:e_2body} we show the evolution of the orbital
eccentricity of two orbiting objects.  The initially circular orbit
picks up a considerable eccentricity due to the copious mass loss in
the first $\sim 10$\,Myr.

At later time, for several simulations around 20\,Myr, the orbital
eccentricity reduces again.  This is a conseqence of mass loss in an
eccentric orbit.  If mass is lost near apocenter the orbit tends to
become less eccentric whereas mass loss at pericenter tends to
increase the eccentricity.  An orbit with a smaller initial mass ratio
tends to become somewhat more eccentric than for a large mass ratio.

Our assumption that mass is lost adiabatically, clearly, is not
supported by the 2-body simulations (see
fig.\,\ref{fig:e_2body}). According to that hypothesis the orbit
should remain circular. But mass in the first few 10\,Myr is lost at
such a high rate that it should in part be treated as an impulsive
event.

Impulsive mass loss is a two body system tends to reduce the absolute
value of the potential energy and the orbital kinetic energy by
decreasing the mass of the system.  Such sudden mass loss imposes no
torgue on the system, so the angular momentum per unit mass remains
the same, but the total angular momentum is reduces as a result of the
decrease in the mass.  Following \citep{1983ApJ...267..322H} we can
then calculate the orbital eccentricity if the total mass of the
binary cluster drops from $M_i$ to $M_f$.  Assuming that the initial
orbit is ciruclar $e_i=0$ the final eccentricity becomes
\citep{1983ApJ...267..322H}.
\begin{equation}
  e_f = { 1 - {M_f/M_i} 
          \over 
        {M_f/M_i} }.
\label{Eq:ecc}
\end{equation}

\section{The $N$-body Simulations}\label{sec:simulation}

We now focus on the $N$-body simulations of binary star clusters.
With the analytic analysis in \S\,\ref{Sect:toymodel} and by
integrating the 2-body systems in \S\,\ref{Sect:2body} we have
acquired some qualitative understanding of the range of initial
conditions to search in order to reproduce the observed cluster pair
NGC\,2136/NGC\,2137.

\subsection{Setting up the realistic simulations}

The main problem in generating proper initial conditions for
NGC\,2136/NGC\,2137 is that we only approximately know the conditions
at an age of about 100\,Myr. In our simulations we will attempt to
start with reasonable initial conditions and see what effect small
changes to those have on the final configuration, and whether or not
the final system has any resemblance with the observed cluster pair.
We explain here how to obtain such initial conditions.

Assuming an initial mass function (Salpeter), age (80--120\,Myr) and
metallicity the mass of the primary cluster (NGC\,2136) can be
estimated from its observed luminosity (B = 10.99, V = 10.7) and turns
out to be about $M =
26300$--28200\,\msun\,\citep{2003MNRAS.338..120M}.  The secondary
cluster (NGC\,2137) is then with $B = 12.66$, $V = 12.88$ about $m =
4500$\,\msun, i.e. the mass ratio $q \equiv m/M \simeq 0.167$.  At a
distance of about 50\,kpc and with a separation between the two
clusters of 1'.34, the current projected separation between the two
cluster is $R \simeq 20$\,pc. Assuming that their orbit is circular we
can compute the tidal radii of the two clusters by estimating the size
of their respective Roche radii with the approximated equation of
\cite{1983ApJ...268..368E}. With the adopted parameters this result in
a tidal radius for the primary cluster of $R_t \simeq 15.0$\,pc, and
for the secondary $r_t \simeq 5.0$\,pc.  The observed core radius of
the primary cluster $R_c \simeq 2.0$\,pc \citep{2003MNRAS.338..120M}.
It turns out that if we assume that the tidal radius for the primary
cluster (NGC\,2136) equals to its Roche radius, it has an unusually
shallow density profile which cannot be described by a
\cite{1966AJ.....71...64K} model or a \cite{1911MNRAS..71..460P}
sphere. During the cluster lifetime its structure is likely to have
changed quite substantially and we are unable to determine the initial
density profile.  For consistency with \S\,\ref{Sect:Makino} we adopt
non-rotating King $W_0=7$ density profiles as the initial conditions
for each of the two clusters in a binary.

We set up the simulations by determining the orbit of the two
clusters. For clarity we adopt here model A\_20 as an example, initial
conditions for the other simulations are presented in
Tab.\,\ref{Tab:Results}.  We start by generating parameters for two
point masses ($M_i = 27500$\,\msun\, and $M_i = 4400$\,\msun, for the
primary and secondary cluster, respectively) in a circular orbit with
a orbital separation of $R_i = 20$\,pc. Later we replace the point
masses with the two clusters.  Each of the clusters is generated by
selecting a number of stars $N_i=9000$ for the primary and $n_i=1500$
for the secondary cluster, both of which are sprinkled in a King
(1966) model with $W_0 = 7$ and a virial radius of $\rvir \simeq
2.14$\,pc for the primary and $\rvir \simeq 0.71$\,pc for the
secondary cluster. With these parameters both clusters are initially
precisely filling their respective Roche lobes.  We subsequently
assign a random mass to each of the stars from a Salpeter initial mass
function between $m_- = 1$\,\msun\, and 100\,\msun, leading to a total
mass of about $M \simeq 27500$\,\msun\, for the primary cluster and $m
\simeq 4400$\,\msun\, for the secondary. We call this models
A\_20. Additional simulations are performed with larger total initial
mass to compensate the mass loss from the evolving stellar
populations. These other models (called B, C and D) are computed with
a primary cluster mass of $M_i \simeq 40800$ and $m_i \simeq 6800$. We
repeated these simulations with $R_i$ varying from 10 to 20\,pc, and
we vary the minimum mass of the initial mass function from $m_- =
1.0$\,\msun\, (model B), $m_- = 0.63$\,\msun\, (model C) to $m_- =
0.50$\,\msun\, (model D).  To study the effect of the initial mass
ratio we performed several additional simulations with $q=0.3$,
$q=0.6$ and $q=1$. The initial conditions are summarized in
Tab.\,\ref{Tab:Results}.

\begin{table*}
\caption[]{
  Initial conditions for the simulations.  The first column gives the
  model name, followed by the initial conditions; initial orbital
  separation ($R_i$), minimum mass of the Salpeter mass function
  ($m_-$), number of stars in the primary ($N_i$) and secondary
  clusters ($n_i$) and the virial radii of the primary ($R_{\rm vir}$)
  and secondary clusters ($r_{\rm vir}$).  The next set of columns
  give the final parameters, time at which we stopped the simulation
  ($t_f$), the number of stars belonging to the primary cluster $N_f$
  and secondary cluster $n_f$ and the distance between the clusters
  $R_f$.  Since none of the A\_10 models survive the last two columns
  give the moment of merger and the number of stars in the merger.
  All simulations ware performed with King models $W_0 = 7$.  The
  simulation naming is chosen as follows: model A, B, C and D identify
  the minimum mass for the initial mass function, followed by the
  initial distance in parsec. All models are computed with a mass
  ratio of $q = 0.167$, except if noted otherwise in the model name,
  in which case the number after the Q identifies the adopted initial
  mass ratio. 
}
\bigskip
\begin{tabular}{lrrrrr|rrrrrr}
Simulation&$R_i$&$m_-$ & $N_i$&$n_i$&$R_{\rm vir}$&$r_{\rm vir}$
                                                 &$t_f$&$N_f$&$n_f$&$R_f$ \\
          &&[\msun]&           &\multicolumn{3}{c}{[pc]}& [Myr]&&& [pc] & \\
\hline
A\_10    &10  &1.0 & 9000& 1500&1.09&0.35& 54&\multicolumn{2}{c}{10499} \\
A\_10Q03 &10  &1.0 & 7875& 2625&1.01&0.41& 50&\multicolumn{2}{c}{10466} \\
A\_10Q06 &10  &1.0 & 6300& 4200&0.94&0.48& 60&\multicolumn{2}{c}{10441} \\
A\_12    &12.5&1.0 & 9000& 1500&1.35&0.43& 90& 9449 &  932&    11.2   \\
A\_15    &15  &1.0 & 9000& 1500&1.63&0.52& 82& 9286 & 1214&    18.4   \\
A\_15Q03 &15  &1.0 & 7875& 2625&1.51&0.62& 96& 8093 & 2310&    19.5   \\
A\_17    &17.5&1.0 & 9000& 1500&1.98&0.60& 81& 8961 & 1539&    25.5  \\
A\_20    &20  &1.0 & 9000& 1500&2.14&0.71& 82& 8948 & 1552&    54.6   \\
A\_20Q03 &20  &1.0 & 7875& 2625&2.02&0.83& 87& 7840 & 2655&    32.4  \\
A\_20Q06 &20  &1.0 & 6300& 4200&1.88&0.97&107& 6215 & 4211&    23.5  \\
A\_20Q10 &20  &1.0 & 5250& 5250&1.42&1.42&117& 5218 & 5207&    26.6  \\
B\_20    &20  &1.0 &13285& 2215&2.14&0.71& 51&13228 & 2272&    32.5   \\
C\_15    &15  &0.63&19543& 3257&1.62&0.51&100&20052 & 2566&    16.9   \\
C\_20    &20  &0.63&19543& 3257&2.14&0.71&100&19703 & 2978&    29.5   \\
D\_20    &20  &0.50&24855& 4145&2.14&0.71& 73&25039 & 3911&    23.1   \\
\hline
\label{Tab:Results}
\end{tabular}
\end{table*}

\subsection{The evolution of simulation models A}\label{Sect:A}

\begin{figure}
\psfig{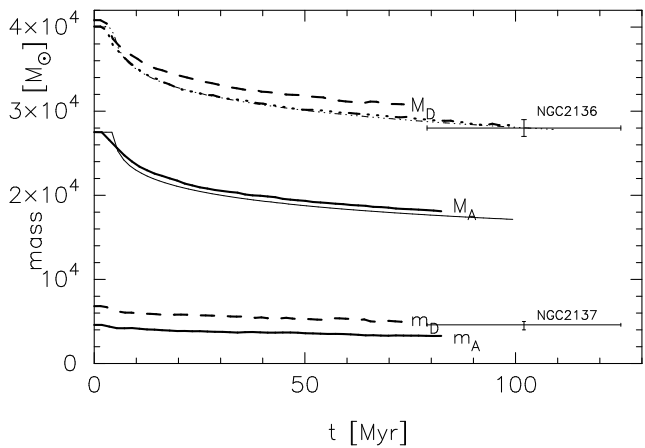}
   \caption[]{
   Evolution of the mass for the primary and secondary clusters in
   models A\_20 (thick solid curves) and for models D\_20 (thick
   dashes) and for the primary cluster in simulation model C\_20
   (thick dash-3-dots).  The error bars to the right indicate the
   beste estimate for the masses of clusters NGC2136 and NGC2137 from
   \cite{2000A&A...360..133D}.  Over-plotted with a thin solid and
   thin dash-3-dotted line are the result of our semi-analytical
   calculation using two point masses for model A\_20 and C\_20. The
   mass loss rate (from \S\,\ref{Sect:toymodel}) matches the full
   N-body simulations quite well and the resulting orbital variation
   is presented in fig.\,\ref{fig:modelABC_Revolution}.
\label{fig:modelA_Mevolution} }
\end{figure}

We will now discuss the results for model A, and continue discussing
models B, C and D in \S\,\ref{Sect:BC}.  The mass lost by the primary
and secondary clusters in model A\_20 and D\_20 are presented in
fig.\,\ref{fig:modelA_Mevolution}. For model C\_20 we only present the
result for the primary cluster.  Over-plotted with thin curves are the
results of a semi-analytic calculation presented in
fig.\,\ref{fig:M_analytic}.  The two thin curves in
fig.\,\ref{fig:modelA_Mevolution} follow the evolution of simulation
models A\_20 (and C\_20) quite satisfactorily.

Mass loss from both clusters in the first few million years is
negligible as only the most massive stars lose mass on the main
sequence. After that the cluster loses mass rapidly, and its effect
becomes quit important. By about 20\,Myr the clusters have lost about
20\% of their mass. The cluster mass loss closely follows the
semi-analytic description we presented in
fig.\,\ref{fig:M_analytic}. Apparently little mass is lost in the form
of escaping stars, consistent with the earlier simulations presented
in \S\,\ref{Sect:Makino}.

As we discussed in \S\,\ref{Sect:analytic} the copious mass loss in
the first few tens of million years drives an expansion of the orbit
and, since the time scale for mass loss and the orbital time scale are
comparable the orbital eccentricity increases.

In simulation A\_20, for example the combined effect of the
adiabatic/impulsive mass loss induces an eccentricity of about $e
\simeq 0.25$ in the first $\sim 20$\,Myr, and the fact that the
cluster approaches apocenter causes the distance between the two
clusters to increase from 20\,pc initially to $\sim 31$\,pc at $t
\simeq 20$\,Myr.  The induced eccentricity on the orbit is consistent
with Eq.\,\ref{Eq:ecc} (see also fig.\,\ref{fig:e_2body}), and also
the evolution of the orbital separation is consistent with the 2-body
simulations in \S\,\,\ref{Sect:2body}.

\begin{figure}
\psfig{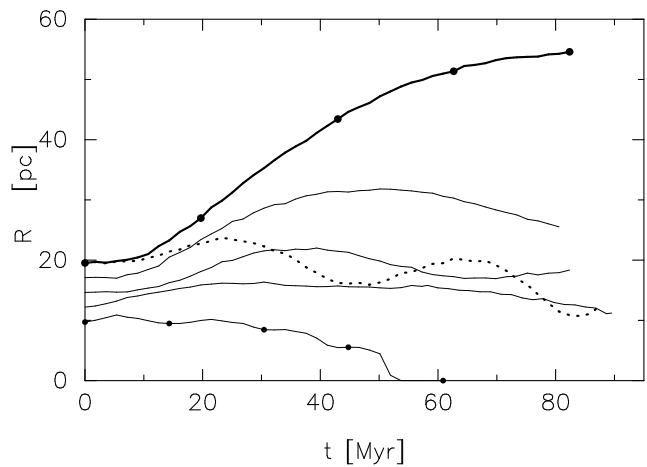}
   \caption[]{
   Evolution of the separation for simulation models A with
   various initial orbital separations (solid curves).  The two curves
   for model A\_20 are for the simulation model without stellar mass
   loss (dotted curve) and with stellar mass loss.  All other curves
   include stellar mass-loss, but start at a different initial
   separation, varying between 10\,pc and 20\,pc. The bullets in the
   top and bottom solid curves indicate the moments in time at which a
   snapshot is presented in Fig.\,\ref{fig:snapshots_modelA}
\label{fig:modelA_Revolution} }
\end{figure}

\begin{figure*}
~\psfig{figure=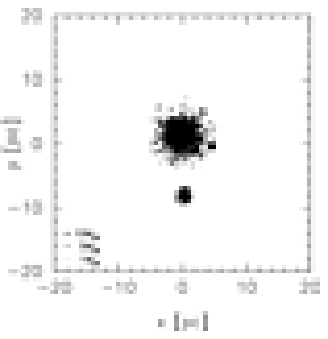,width=0.4\columnwidth,angle=-90}
~\psfig{figure=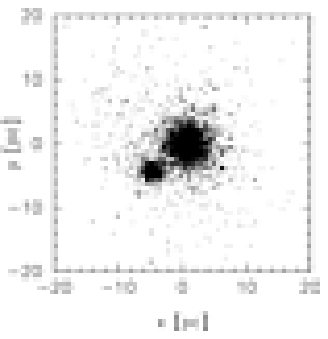,width=0.4\columnwidth,angle=-90}
~\psfig{figure=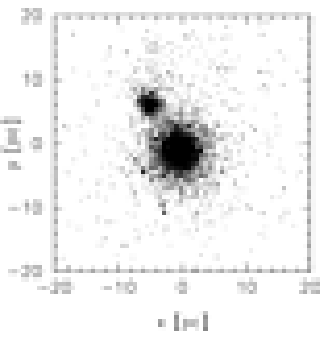,width=0.4\columnwidth,angle=-90}
~\psfig{figure=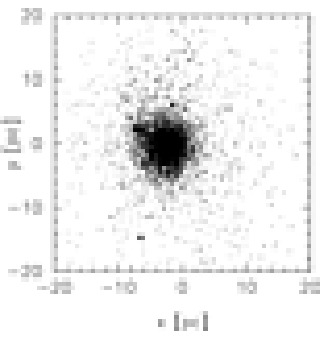,width=0.4\columnwidth,angle=-90}
~\psfig{figure=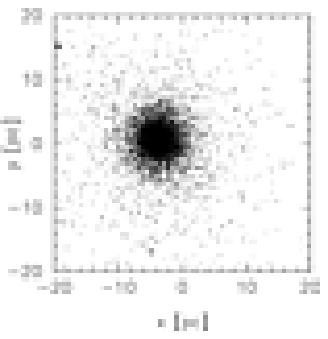,width=0.4\columnwidth,angle=-90}
~\psfig{figure=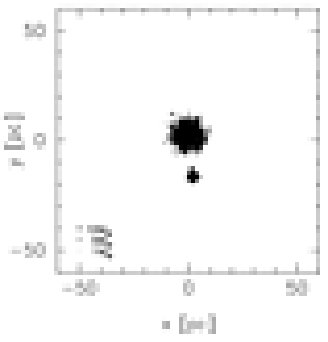,width=0.4\columnwidth,angle=-90}
~\psfig{figure=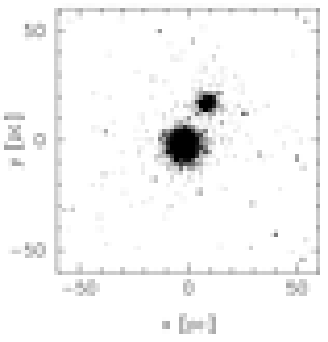,width=0.4\columnwidth,angle=-90}
~\psfig{figure=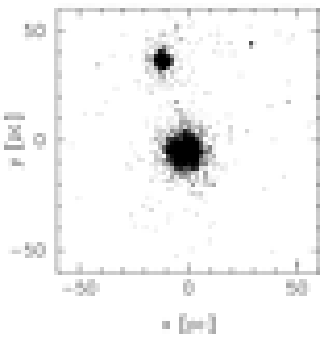,width=0.4\columnwidth,angle=-90}
~\psfig{figure=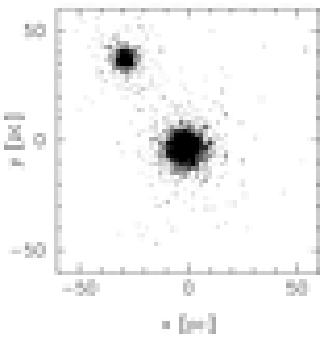,width=0.4\columnwidth,angle=-90}
~\psfig{figure=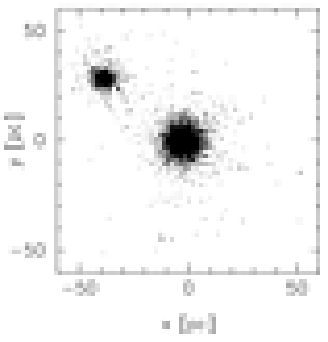,width=0.4\columnwidth,angle=-90}
\caption[]{
  Representation in the $X$--$Y$-plane for models A\_10 (top panels)
  and A\_20 (bottom panels).  The simulations models A\_10 are given
  at birth (left image), at an age of about 15\,Myr, 30\,Myr, 45\,Myr
  and at 60\,Myr (right), whereas for simulation model A\_20 the
  images represent the cluster birth (left image), at an age of
  21\,Myr, 42\,Myr, 63\,Myr and at 84\,Myr (right).  The moments at
  which snapshots were taken are identified with bullets in
  Fig.\,\ref{fig:modelA_Revolution}.  ({\em Note concerning this LANL
  version: the published version will contain high quality images.})
\label{fig:snapshots_modelA} }
\end{figure*}

In fig.\,\ref{fig:snapshots_modelA} we present a representation of the
stellar positions for simulation model A\_10 (top panels) and A\_20
(bottom) at various moments in time.  We stop both simulations when they
reach an age of at least $\sim 80$\,Myr.  The three cluster pairs of model
A\_10 have merged before that time, whereas the distance between the
two clusters of model A\_20 has then increased beyond 50\,pc.

The simulations in which the clusters are initially rather widely
separated can satisfactorily be described by the semi-analytic
analysis in \S\,\ref{Sect:analytic}. In close proximity the transport
of angular momentum from the orbit to the internal rotation becomes
gradually more important (see \S\,\ref{Sect:Makino}): tighter cluster
binaries are able to transport angular momentum more effectively from
the orbit to the rotation of the individual clusters, but also to
stellar escapers.  These dynamical interferences dramatically reduces
the predictability of a simple semi-analytic model, because it is not
apriory clear what amount of specific angular momentum in stored in
the cluster. For close pairs, stellar mass loss appears to be
relatively less important whereas the redistribution and loss of
angular momentum dominates the evolution of the orbital parameters.
This is illustrated in Fig.\,\ref{fig:modelA_Revolution}, where we
present the evolution of the orbital separation for simulation models
A with an initial separation of 20\,pc, 17.5\,pc, 15\,pc, 12.5\,pc and
10\,pc (solid curves from top to bottom).  The orbits of the clusters
at 10\,pc, 12.5\,pc and 15\,pc expand less dramatically than expected
on the amount of mass lost from the evolving stellar populations.

The separation between the two clusters in models A\_15 increases in
the first $\sim 40$\,Myr by mass loss and the orbital eccentricity
increases but the loss of angular momentum prevent that the clusters
continue to recede, as is the case in model A\_20. In the latter model
the two clusters approach periclustron again at about 140\,Myr.
Simulations A\_12 to A\_17 approach apoclustron at an age of about
$\sim 40$\,Myr, afther which their separations decreases again. These
clusters can survive for a long time as binaries as they spiral-in
only rather slowly.  Once stellar evolution becomes less important (at
a turn-off mass of $\sim 8$\,\msun\, i.e; after about 40\,Myr) the
dynamical redistribution of angular momentum becomes dominant again in
driving the orbital evolution of the binary.

In simulation model A\_10 (lower solid curve in
Fig.\,\ref{fig:modelA_Revolution}) the stellar mass loss cannot
compete with the loss of orbital angular momentum.  The two clusters
merge as soon as the gyro-dynamical instability sets in, which happens
as soon as the gyration radius of the two clusters exceeds the orbital
separation (see Eq.17 of \cite{1991Ap&SS.185...63M}). This happens at
about 50\,Myr.  Stellar evolution in this case delays the merger by
about a factor of two, but cannot prevent it.  The gyration radius of
a self gravitating system of stars can to first order be approximated
with $r_{\rm vir}$.

Clusters with an initial separation between about 12 and 18\,pc, have
the best change to survive to an age of $\sim 100$\,Myr.  In this
range of initial separations the orbital evolution is strongly
affected by stellar mass loss, but also by the redistribution of
angular momentum, making this regime the most interesting to study
numerically.  If the orbital separation exceeds about 20--30\,pc the
background tidal field starts to perturb the orbital evolution of the
cluster binary. In addition, clusters with such wide separations are also
vulnerable to being ionized by passing giant molecular clouds, making
such wide binaries unlikely to survive for an extensive period of
time.

In the range in orbital separation between about 12\,pc and 18\,pc both
clusters fill their respective Roche lobes throughout the simulation,
and stars stream through the first Lagrangian point from the secondary
cluster to the primary, and vice versa.  Note that few stars are lost
through the other Lagrangian points and we find no evidence for tidal
tails in any of our simulations, contrary to what is observed in some
apparent binary clusters in the LMC \citep{1999A&A...344..450L}.

\begin{figure}
\psfig{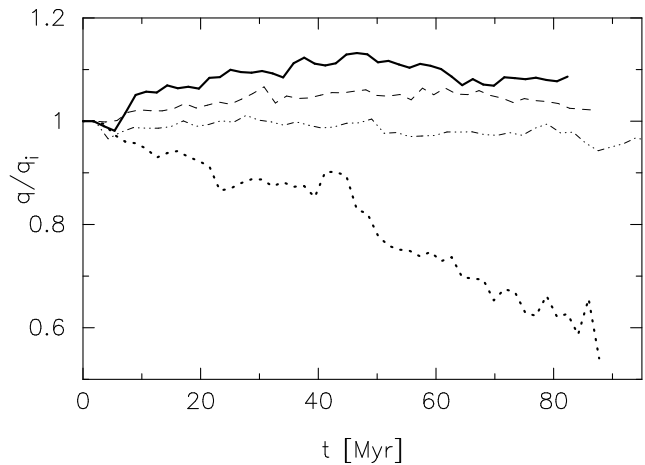}
   \caption[]{
   Evolution of the mass ratio ($q \equiv m/M$) normalized to the
   initial mass ratio for simulation model A\_20.  The solid and
   dotted curves are for an initial mass ratio $q_i = 0.167$.  For the
   solid curve stellar mass loss was taken into account, whereas the
   dotted curve presents the simulation in which stellar evolution was
   ignored.  The dashed and dash-3-dotted curves are for $q_i = 0.33$
   and $q_i = 0.67$, respectively. The evolution of the mass ratio for
   simulation model A\_20 with $q_i = 1$ is statistically
   indistinguishable for the dash-3-dotted curve.
\label{fig:modelA_Qevolution} }
\end{figure}

The number of stars transferred from one cluster to the other is
generally small.  By the end of the simulations a total of $238\pm42$
stars that are originally born in the primary cluster end up as
members of the secondary cluster, whereas $517\pm302$ stars from the
secondary cluster are found in the primary cluster (the error here is
estimated from the dispersion between the different runs). The mass
ratio of the binary remained roughly constant with time, as is
illustrated in fig.\,\ref{fig:modelA_Qevolution}.  In this figure we
present the evolution of the mass ratio for simulation model A\_20
with various initial mass ratios.  Also in the other simulations the
mass ratio hardly changes with time (see Tab.\,\ref{Tab:Results}).
The simulation without stellar mass loss (dotted curve in
fig.\,\ref{fig:modelA_Qevolution}) experiences the largest variation
in mass ratio, as in this model stars effectively stream from the
secondary to the deeper potential well of the primary cluster (see
also \cite{1991Ap&SS.185...63M}).

\begin{figure}
\psfig{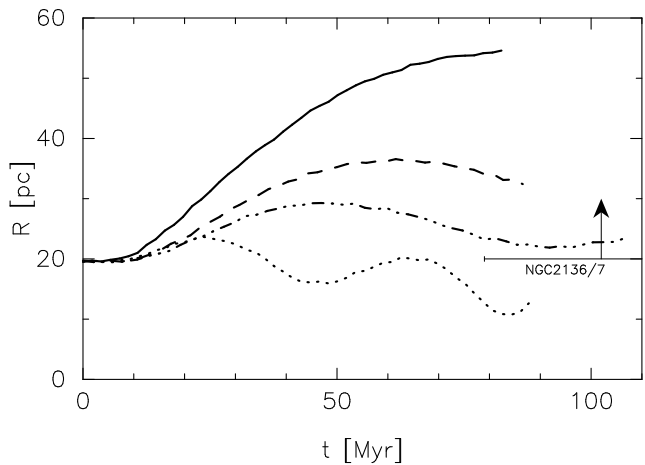}
   \caption[]{
   Evolution of the separation for simulation models A at an initial
   separation of 20\,pc for various choice of the initial mass ratio.
   The solid and dotted curves are for the simulation models A\_20
   with a mass ratio of 0.167 while including stellar mass loss (solid
   curve) and without stellar mass loss (dotted curve).  The dashed
   and dash-3-dotted curves are for a mass ratio of 0.33 and 0.66,
   respectively, while taking the total number of stars the same.
\label{fig:modelA_Revolution_varQ} }
\end{figure}

For completeness, in fig.\,\ref{fig:modelA_Revolution_varQ} we present
the evolution of the orbital separation for some of the simulation
models presented in fig.\,\ref{fig:modelA_Qevolution}.

The behavior of the cluster binary is substantially affected by
stellar mass loss, and the lower limit of the initial mass function is
therefore an important parameter (see \S\,\ref{Sect:analytic}). Mass
loss in the early stage is governed by the choise of $m_-$ and the
slope of the IMF. In the next section we will further explore the
range in initial cluster separations for which stable binaries exist
by varying the choise of $m_-$.

\subsection{The evolution of simulation models B, C and D}\label{Sect:BC}

Mass loss in the early evolution of binary star clusters is of major
importance (see figs.\,\ref{fig:a_analytic}, \ref{fig:a_2body},
\ref{fig:modelA_Mevolution} and \ref{fig:modelA_Revolution}). Most
mass is lost by the evolving stellar population and a different
initial mass function may therefore have dramatic consequences to the
orbital evolution of the binary.  One of the effects of mass loss is
that the simulated clusters in model A were substantially less massive
than observed.  To compensate for this we increase the initial mass of
the primary cluster from 27500\,\msun\, to 40800\,\msun.  Since the
mass ratio for the more realistic models remains roughly constant with
time (see fig.\,\ref{fig:modelA_Qevolution}) we keep the same mass
ratio of $q = 0.167$.

Much of the mass loss behavior in our simulation models is dominated
by the choise of the lower limit to the initial mass function.  Our
adopted lower limit of $m_- = 1$\,\msun\, can be considered on the
high side (see however
\cite{2001MNRAS.326.1027S,2002ApJ...581..258F,2005ApJ...628L.113S}).
We study the effect of reducing $m_-$ while increasing the number of
stars to keep the total mass constant, but keeping all other
parameters as much as possible the same.  For a minimum mass $m_- =
0.63$\,\msun\, we requires a total of 22800 stars, and for $m_-=
0.50$\,\msun\, the total number of stars in the simulation increases
to $29000$. For practical reasons we do not simulate larger clusters,
though in principle it is possible to redo these simulations with a
\cite{2001MNRAS.322..231K} initial mass fucntion with the hydrogen
burning limit as a minimum mass. This would increase the number of
stars in our simulation to $\simeq 70000$, which is beyond the scope
of the current paper.

\begin{figure}
\psfig{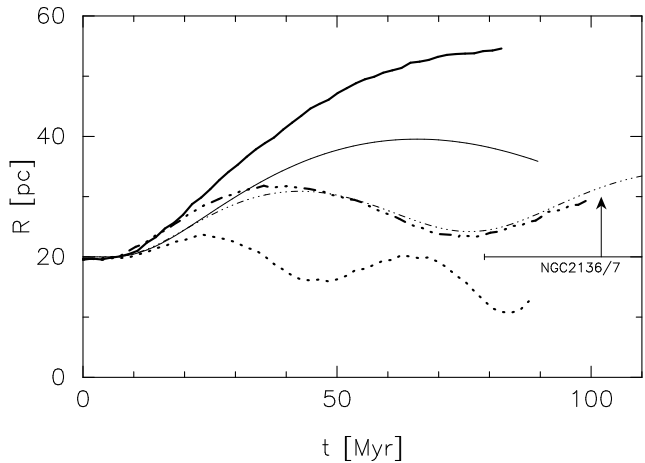}
   \caption[]{
   Evolution of the separation for simulation models at an initial
   separation of 20\,pc.  The top (thick solid) curve is for
   simulation model A\_20, the middle (thick dash-3-dotted) curve is
   for model C\_20.  The bottom (dotted) curve gives the evolution of
   the orbital separation for model A\_20 but without stellar mass
   loss.
   The thin solid and dash-3-dotted curves give the results of our
   2-body simulations of \S\,\ref{Sect:2body} with identical initial
   conditions as for models A\_20 and C\_20\, respectively.  The mass
   loss for these models are presented in
   fig.\,\ref{fig:modelA_Mevolution}.
   The observed projected separation between NGC2136 and NGC2137 is
   printed to the right as a lower limit.
\label{fig:modelABC_Revolution} }
\end{figure}

The evolution of the mass of the primary cluster for models C\_20 and
D\_20 are presented in fig.\,\ref{fig:modelA_Mevolution}, and both
match the observed mass of NGC\,2136 and NGC\,2137. We present the
evolution of the orbital separation for these runs in
fig.\,\ref{fig:modelABC_Revolution}.  For comparison we also show
models A\_20 with and without stellar mass loss.  Due to the larger
proportion of low mass stars in simulation model C\_20 and D\_20
stellar mass loss is less important than in model A\_20. As a
consequence the orbit (and clusters) expans less dramatically.  In
addition, the increased number of stars, and therefore the number
density in the cluster, allows for angular momentum to be carried away
by escapers more effectively. The result is a less pronounced increase
in the orbital separation compared to model A\_20 (see
\S\,\ref{Sect:A}). The induced orbital eccentricity is also smaller.
For comparison with the observations we over-plot the projected
distance between NGC\,2136 and NGC\,2137 in
Fig.\,\ref{fig:modelABC_Revolution}.

Over-plotted with the thin solid and dash-3-dotted curves in
fig.\,\ref{fig:modelABC_Revolution} are the result of orbital
expansion by solving the two-body problem which includes stellar mass
loss (see \S\,\ref{Sect:2body}).  The lower (thin) solid curve gives
the orbial evolution of the 2-body results of model A\_20, whereas the
thick curve gives the results of the full $N$-body simulation of the
same model. We also show the 2-body and the full $N$-body results for
model C\_20 (see the thin and thick dash-3-dotted curves,
respectively).  To illustrate the effect of the redistribution of
angular momentum we also show the result of the full $N$-body
simulation of model A\_20 but without stellar mass loss (dotted
curve).

One expects naively that the $2$-body integration should
systematically overestimate the orbital expansion, but for model
A\_20, this is clearly not the case, as here the 2-body model falls
short by about 40\% with respect to the $N$-body simulation (thick
solid curve).  The orbital expansion of simulation model C\_20,
however, is quite well reproduced with the 2-body solution.  Remember
here, however, that the semi-analytic models are rather simplistic, as
the 2-body problem is solved without accouting for the drag and
friction between the clusters and, ignores the possibility of the
redistribution of rotational angular moment of the clusters and the
orbital angular momentum, whereas Eq.\,\ref{Eq:afai} gives only an
orbital average result.

The best match between the simulations and the observed clusters is
obtained with an initial separation of about 20\,pc with an initial
mass function that extends to lower masses (0.5--0.6\,\msun).  For
these models, also the total mass and the mass ratio of the clusters
at an age of 100\,Myr matches the observations quite
satisfactorily. For model C\_15 the orbital separation at an age of
100\,Myr falls short of the observed separation.  We therefore
conclude that models C\_20 and D\_20 most satisfactorily reproduce the
characteristics of the oberved cluster pair NGC\,2136/NGC2137.

\subsection{Evolution of the cluster structure}

\begin{figure}
   \psfig{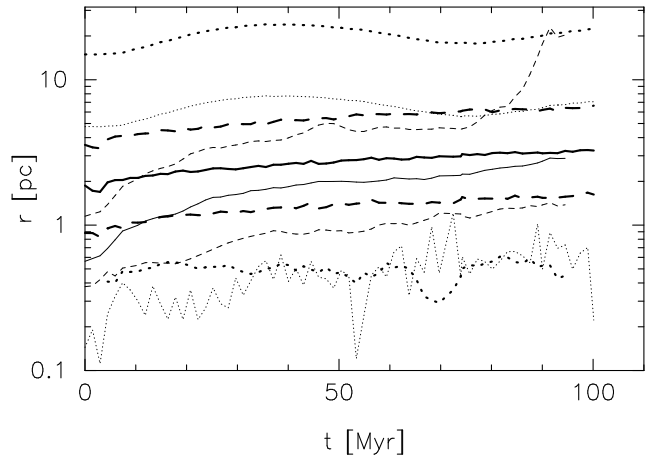}
   \caption[]{
   Evolution of the Lagrangian radii for simulation models C\_20. The
   thick curves are for the primary cluster whereas the thin lines are
   for the secondary cluster.  From bottom to top the curves are core
   radius (dotted curves), 25\% (dashes), 50\% (solids) and 75\%
   (upper dashes) Lagrangian radii.  The top thick and thin dotted
   curves give the tidal radius for the primary and secondary cluster,
   respectively. Note that we smoothed the data for the secondary
   cluster over a co-moving window using 4 points spanning about
   6\,Myr.
\label{fig:modelB_Lagrangian_radii} }
\end{figure}

In figure\,\ref{fig:modelB_Lagrangian_radii} we present the evolution
of the Lagrangian radii for simulation model C\_20.  Stellar mass loss
has a dramatic effect on the evolution of the internal structure of
the clusters. In particular the secondary cluster (thin lines in
fig.\,\ref{fig:modelB_Lagrangian_radii}) experiences a dramatic
expansion in the first several 10\,Myr due to stellar mass loss and
internal dynamical evolution.  Core collapse in this cluster occurs at
about 2\,Myr, for the primary cluster a shallow core collapse is
reached at about 70\,Myr.  Due to the induced rotation by their mutual
tidal coupling, the core collaps in both clusters may happen somewhat
earlier than naively expected for isolated clusters.  Rotation can
initiate a gravogyro catastrophe through the transport of angular
momentum \citep{1989PASJ...41..991A,1999ASPC..182..105S}.

The post collapse expansion of the secondary cluster exceeds the
adiabatic expansion driven by stellar mass loss.  Since the secondary
cluster expands more quickly than the orbital separation it continues
to over-fill its Roche lobe and transfer mass to the primary cluster,
whereas the primary cluster detaches from its Roche lobe. As a result
the orbital separation between 10\,Myr and 40\,Myr increases even more
dramatically than naively expected from the rapid mass loss by stellar
evolution alone (see fig.\,\ref{fig:modelABC_Revolution}). It is
interesting to note that the evolution of a binary cluster is governed
by the more rapid dynamical evolution of the secondary cluster,
contrary to stellar binaries in which it is generally the primary star
which evolves first.

\section{Discussion}\label{sec:Discussion}

We studied the evolution of bound pairs of star clusters by means of
direct N-body simulations while including the effects of stellar mass
loss.  The initial parameters selected for our study are loosely based
on the binary star cluster NGC\,2136 and NGC\,2137 in the large
Magellanic cloud.

When starting with an initial cluster mass of about 40800\,\msun\, and
6800\,\msun\, in a circular orbit with a separation of 15--20\,pc our
simulations at about 100\,Myr are consistent with the masses and
orbital separation of the observed cluster pair.  The observations are
reproduced most satisfactorily when we adopt a Salpeter mass function
with a lower mass limit of 0.5--0.6\,\msun, but lower mass cut-offs
may also produce a satisfactory comparison with the observations.

A word of caution, however, is well placed here, as we did not cover
the entire parameters space in excruciating detail. The initial
density profile, not varied in our study, may have a profound effect
on the evolution of the binary cluster. The main aim of this paper is
therefore not to constrain the initial conditions for
NGC\,2136/NGC\,2137, but rather to obtain a better understanding of
the general evolution of binary star clusters. Earlier numerical
studies of binary star clusters ignored the stellar mass spectrum,
stellar evolution and used approximate N-body techniques
\citep{1989PASJ...41.1117S,1991Ap&SS.185...63M,1998MNRAS.295..921D}.
In our simulations all these effects are included in a self consistent
fashion.

We are pleasantly surprised by the rich dynamics embedded in the
evolution of binary star clusters. In particular the finding that
stellar mass loss in the early evolution of the cluster is rapid
enough to be considered a shock for the orbital elements of the binary
cluster came somewhat unexpected. The consequence is that the orbits
of star clusters with an age $\apgt 10$\,Myr are not circular, but
small eccentricities $e\apgt 0.2$ are induced upon the orbit, even if
the two clusters are in Roche-lobe contact. The evolution of the
distance between the two clusters is than a consequence of the orbital
evolution (adiabatic/impulsive expansion) and the fact that cluster
pair aproaches apoclustron.  Note, however, that all our models
started with circular orbits. This choise was to limit parameters
space as introducing an initial eccentricity also requires the choise
of an eccentric anomaly.

Our initial clusters are not rotating. Since the redistribution of
angular momentum turns out to be a mayor effect in the evolution of
the cluster binary, future simulation may study the evolution of star
cluster binaries by taking this extra parameter into account.

Another interesting aspect of our study that the less massive cluster
is more likely to overfill its Roche lobe. The secondary cluster
experiences core collapse in an earlier stage than the primary
clusters. The resulting expansion of the post collapsed secondary
cluster subsequently drives the transfer of mass through the first
Lagrangian point to the primary cluster. Mass transfer proceeds from
the less massive to the more massive component but the total number of
stars transferred is relatively small (1--3\%). The orbital evolution
is dominated by stellar mass loss and by redistribution of angular
momentum and by escaping stars

During the exchange of stars from one cluster to the other hardly any
stars are lost through the second and third Lagrangian points. Stars do
escape the clusters through the first Lagrangian point and
isotropically. The isotropic escapers are mainly neutron stars which
often receive high kick velocities in a supernova explosion. We
therefore see no indication that tidal tails form in binary
clusters. The origin of the tidal tails in several observed multiple
clusters in the small Magellanic cloud by \cite{1999A&A...344..450L}
and \cite{2000A&AS..146...57D} is therefore unlikely to be caused by 
the binarity of these clusters.

Binary star clusters in the LMC with a separation exceeding about
20\,pc are vulnerable to ionization by giant molecular clouds. Our
initial conditions approach this boundary. Such widely separated
clusters can survive because the encounter rate between clusters in
the LMC is rather small, making it unlikely that a cluster binary is
ionized within a few times 10\,Myr. However, as we demonstrated with
our simulations, the orbital separation may increase considerably,
making such clusters much more vulnerable to close (ionizing)
encounters with other clusters or giant molecular clouds. Ionization
by the background tidal field of the LMC is also possible. We did not
take these effects into account in our simulations, but the
consequences may be profound. One consequence is that there are (at
least) two reasons why binary clusters in older populations are rare:
close cluster pairs tend to merge and wide cluster pairs are likely to
be ionized by the background tidal field or by passing Molecular
clouds.  Of course, we cannot exclude that binary clusters are simply
rarely formed.

If the distance between the two clusters is initially sufficiently
large that the redistribution of angular momentum and
escapers has little effect on the orbital evolution, semi-analytic
calculations about the orbital evolution of the binary cluster produce
satisfactory results. As soon as dynamical effects start to become
important (at separations $\aplt 15$\,pc) these approximations break
down.

\section*{Acknowledgments}

The first version of this paper was written in 'Taylor house' and SPZ
is grateful to Helen Thompson-Taylor for her wonderful hospitality. We
are grateful to Alessia Gualandris, Thijs Kouwenhoven, Jun Makino and
Steve McMillan for many discussions.  SPR thanks the Astronomical
institute 'Anton Pannekoek' and the Section Computational Science of
the University of Amsterdam for their hospitality.  This research was
supported in part by the Netherlands Organization for Scientific
Research (NWO grant No. 635.000.001 and 643.200.503), the Netherlands
Advanced School for Astronomy (NOVA), the Royal Netherlands Academy
for Arts and Sciences (KNAW), the Leids Kerkhoven-Bosscha fonds (LKBF)
by NASA ATP grant NNG04GL50G and by the National Science foundation
under Grant No. PHY99-07949.  All calculations are performed using the
MoDeStA GRAPE-6 systems in Amsterdam.

\def\nat{\ {Nat}\ }
\def\aa{\ {A\&A}\ }
\def\aap{\ {A\&A}\ }
\def\aaps{\ {A\&AS}\ }
\def\mnras{\ {MNRAS}\ }
\def\apss{\ {Ap\&SS}\ }
\def\aj{\ {AJ}\ }
\def\apj{\ {ApJ}\ }
\def\apjl{\ {ApJL}\ }
\def\pasj{\ {Publ. Astr. Soc. Japan}\ }

\bsp
\label{lastpage}
\end{document}